\documentclass[aps,twocolumn,pra,showpacs,floatfix]{revtex4}
\usepackage{epsfig}
\usepackage{graphicx}
\usepackage{dcolumn}

\begin{document}

\title{Sensitivity of the energy levels of singly ionized cobalt to 
the variation of the fine structure constant}
\author{V. A. Dzuba and V. V. Flambaum}
\affiliation{School of Physics, University of New South Wales, Sydney 2052, 
Australia}

\date{\today}

\begin{abstract}

We use relativistic Hartree-Fock and configuration interaction methods
to calculate the dependence of transition frequencies for singly
ionized cobalt on the
fine structure constant. The results are to be used in the search for
variation of the fine structure constant in quasar absorption spectra. 

\end{abstract}


\maketitle

\section{Introduction}

Search for variation of fundamental constants is motivated by theories
unifying gravity with other interactions as well as by many
cosmological models. The search spans the whole lifetime of the
universe from Big-Bang nuclear synthesis to the present-day very precise
atomic clock experiments (see,
e.g. reviews~\cite{Uzan,Flambaum07a,Dzuba09,Flamb09}). 
No unambiguous manifestation of the variation of fundamental constants
have been found so far. However, there is large amount of data which
is consistent with variation of the fine structure constant or the
ratio of the electron to proton
mass~\cite{Uzan,Flambaum07a,Dzuba09,Flamb09}).  Most of this data
comes from the analysis of the quasar absorption spectra. The analysis
of the data obtained on the Keck telescope in Hawaii indicate that the
fine structure constant $\alpha$ might be smaller in early
universe~\cite{Webb99,Webb01,Murphy01a,Murphy01b,Murphy01c,Murphy01d}. 
However, an analysis of the data
from the VLT telescope in Chile, performed by different groups~\cite{vlt1,vlt2}
gave a null result. There is an intensive debate
in the literature about possible reasons for the disagreement 
(see. e.g.~\cite{Flambaum07,Srianand07,Murphy08,Griest}).

The most probable reason for disagreement is the effect of some
unknown systematics. One of the ways to deal with any unknown
systematics is to include as many atomic lines into analysis as
possible and compare the results for different lines for
consistency. Atomic frequencies in different atoms and even
frequencies of different transitions in the same atom depend on the
fine structure constant very differently. It is extremely unlikely that
any unknown systematics would behave exactly the same way thus
mimicking the variation of the fine structure constant. 

It was recently brought to our attention that some lines of
single-ionized cobalt are observed in the quasar absorption
spectra~\cite{cobalt}. To include these lines into analysis one needs to
know how the frequencies of the corresponding transitions depend on the
fine structure constant. To reveal this dependence we perform atomic
calculations following the technique developed in our previous
works~\cite{Dzuba08a,Dzuba08b}.  

\section{Method}

\begin{table} 
\caption{Configurations and effective core polarizabilities
  [$\alpha_c$ (a.u.)] used in the calculations.}
\label{Tab:a}
\begin{tabular}{lll}
\hline \hline
\multicolumn{1}{c}{Parity} & 
\multicolumn{1}{c}{Configuration} & 
\multicolumn{1}{c}{$\alpha_c$} \\
\hline
Even & $3d^8$      & 0.4 \\
Even & $3d^74s$    & 0.5177 \\
Odd  & $3d^74p$    & 0.448 \\
Odd  & $3d^64s4p$  & 0.75 \\
\hline \hline
\end{tabular}
\end{table}

The dependence of atomic frequencies on the fine-structure constant
$\alpha=e^2/\hbar c$ appears due to relativistic corrections. In the
vicinity of its physical value $\alpha_0 = 1/137.036$ it is
presented in the form 
\begin{equation}
  \omega(x) = \omega_0 + qx,
\label{omega}
\end{equation}
where $\omega_0$ is the present laboratory value of the frequency and
$x = (\alpha/\alpha_0)^2-1$, and $q$ is the coefficient which is to be
found from atomic calculations. Note that
\begin{equation}
 q = \left .\frac{d\omega}{dx}\right|_{x=0}.
\label{qq}
\end{equation}
To calculate this derivative numerically we use
\begin{equation}
  q \approx  \frac{\omega(x) - \omega(-x)}{2x}.
\label{deriv}
\end{equation}
Here $x$ must be small to exclude non-linear in $\alpha^2$ terms.
In the present calculations we use $x = 0.01$.

The atomic structure calculations are performed with the use of the
relativistic Hartree-Fock method (RHF) and the configuration
interaction technique (CI). The RHF self-consistent procedure is done
separately for two even and two odd configurations presented in
Table~\ref{Tab:a}. The resulting $3d$, $4s$, and $4p$ single-electron
functions are used as a basis for the CI calculations. Note that these
states are different in different configurations. For example the $3d$
state in the $3d^8$ configuration is not the same as the $3d$ state in
the $3d^74p$ configuration, etc. See Ref.~\cite{Dzuba08a} for details.

The effective Hamiltonian for $N_v$ valence electrons has the form

\begin{equation}
  \hat H^{\rm eff} = \sum_{i=1}^{N_v} \hat h_{1i} + 
  \sum_{i < j}^{N_v} e^2/r_{ij},
\label{heff}
\end{equation}
here, $\hat h_1(r_i)$ is the one-electron part of the Hamiltonian:
\begin{equation}
  \hat h_1 = c \mathbf{\alpha \cdot p} + (\beta -1)mc^2 - \frac{Ze^2}{r} 
 + V_{core} + \delta V.
\label{h1}
\end{equation}
Here $\mathbf{\alpha}$ and $\beta$ are Dirac matrixes, $V_{core}$ is
Hartree-Fock potential due to core electrons and $\delta V$
is the term which simulates the effect of the correlations between core
and valence electrons. It is often called the {\em polarization
  potential} and  has the form
\begin{equation}
  \delta V = - \frac{\alpha_c}{2(r^4+a^4)}.
\label{dV}
\end{equation}
Here $\alpha_c$ is the polarization of the core and $a$ is a cutoff
parameter (we use $a = a_B$). We treat $\alpha_c$ as fitting
parameters and choose their values to reproduce experimental position
of the configurations on the energy scale. Corresponding values of
$\alpha_c$ are presented in Table~\ref{Tab:a}.

\section{Results}

\begin{table} 
\caption{Energies (cm$^{-1}$), $g$-factors, and $q$-coefficients for
  odd states of Co, which are accessible from the ground state
  ($^3$F$_4$) via electric dipole transition.}
\label{Tab:q}
\begin{tabular}{llllllr}
\hline \hline
&\multicolumn{2}{c}{Experiment\footnotemark[1]} &
&\multicolumn{3}{c}{Calculations} \\
\multicolumn{1}{c}{State} &
\multicolumn{1}{c}{Energy} &
\multicolumn{1}{c}{$g$} &
\multicolumn{1}{c}{$g_{NR}$} &
\multicolumn{1}{c}{Energy} &
\multicolumn{1}{c}{$g$} &
\multicolumn{1}{c}{$q$} \\
\hline
$z^5$F$^o_3$ & 45972.17 &   1.30  &  1.30   &  46021 &  1.3057 & -872 \\
$z^5$D$^o_3$ & 47039.27 &   1.442 &  1.50   &  46715 &  1.3865 & -444 \\
$z^5$G$^o_3$ & 48151.07 &   0.927 &  0.9167 &  47221 &  0.9498 &   50 \\
$z^3$G$^o_3$ & 50036.55 &   0.80  &  0.75   &  48481 &  0.8572 &  368 \\ 
$z^3$F$^o_3$ & 50381.86 &   1.055 &  1.083  &  48148 &  1.0910 & -277 \\
$z^3$D$^o_3$ & 51512.41\footnotemark[2]
                  &   1.32  &  1.33   &  47606 &  1.2426 &-1067 \\ 
$y^5$D$^o_3$ & 61240.96 &   1.505 &  1.50   &  64844 &  1.4971 &-1054 \\  
$z^5$P$^o_3$ & 63344.50 &   1.67  &  1.67   &  65862 &  1.6504 & -977 \\ 
$y^3$D$^o_3$ & 63587.01\footnotemark[2]
                       &   1.35  &  1.33   &  66075 &  1.3407 & -979 \\ 
$y^3$F$^o_3$ & 64360.26 &   1.06  &  1.08   &  65509 &  1.0577 & -964 \\
$y^3$G$^o_3$ & 65174.89 &   0.78  &  0.75   &  66286 &  0.7615 & -280 \\
$z^1$F$^o_3$ & 66017.79 &   1.01  &  1.00   &  66678 &  1.0111 & -508 \\
$x^3$D$^o_3$ & 67524.20\footnotemark[2]
                        &   1.33  &  1.33   &  71424 &  1.3101 &-1245 \\ 
$w^3$D$^o_3$ & 69060.26 &   1.31  &  1.333  &  73395 &  1.3065 &-1026 \\
$x^3$G$^o_3$ & 69356.64 &   0.78  &  0.750  &  72125 &  0.7578 &  -50 \\
$x^3$F$^o_3$ & 70457.93 &   1.09  &  1.083  &  73985 &  1.0643 & -974 \\

$z^5$F$^o_4$ &  45378.85 &   1.42  &  1.35 &   45393 &  1.4078 & -1400 \\
$z^5$D$^o_4$ &  46320.96 &   1.447 &  1.50 &   45991 &  1.4115 & -1100 \\
$z^5$G$^o_4$ &  47807.58 &   1.154 &  1.15 &   46732 &  1.1564 &  -500 \\
$z^3$G$^o_4$ &  49348.43\footnotemark[2]
                         &   1.11  &  1.05 &   47804 &  1.0850 &  -300 \\
$z^3$F$^o_4$ &  49697.81\footnotemark[2]
                         &   1.19  &  1.25 &   47080 &  1.2376 &  -900 \\
$y^5$D$^o_4$ &  61388.43 &   1.51  &  1.50 &   64773 &  1.4953 & -1000 \\
$y^3$F$^o_4$ &  63510.40\footnotemark[2]
                         &   1.152 &  1.25 &   65074 &  1.2273 & -1311 \\
$z^3$H$^o_4$ &  63792.90\footnotemark[2]
                         &   0.86  &  0.80 &   65490 &  0.9591 &  -957 \\
$z^1$G$^o_4$ &  64401.46 &   1.07  &  1.00 &   65814 &  0.8933 &  -522 \\
$y^3$G$^o_4$ &  65154.18 &   1.06  &  1.05 &   66503 &  1.0185 &  -290 \\
$x^3$G$^o_4$ &  68843.67 &   1.05  &  1.05 &   71639 &  1.0439 &  -715 \\
$x^3$F$^o_4$ &  70186.30\footnotemark[2]
                         &   1.26  &  1.25 &   73782 &  1.2505 &  -962 \\
$y^3$G$^o_4$ &  72009.0  &   0.70  &  0.80 &   73090 &  0.8322 &  -130 \\
$y^1$G$^o_4$ &  73147.23 &   1.04  &  1.00 &   72734 &  0.9816 &  -431 \\

$z^5$F$^o_5$ &  45197.78 &   1.386 &  1.40 &   45078 &  1.3924 & -1568 \\
$z^5$G$^o_5$ &  47345.94 &   1.260 &  1.27 &   46065 &  1.2482 & -1363 \\
$z^3$G$^o_5$ &  48556.16\footnotemark[2]
                         &   1.194 &  1.20 &   46872 &  1.2248 &  -812 \\
$z^3$H$^o_5$ &  63306.66 &   1.03  &  1.03 &   65002 &  1.0312 & -1274 \\
$y^3$G$^o_5$ &  64601.72 &   1.16  &  1.20 &   65453 &  1.1756 & -1151 \\
$z^1$H$^o_5$ &  64957.50 &   1.05  &  1.00 &   66126 &  1.0212 &  -259 \\
$x^3$G$^o_5$ &  68203.39\footnotemark[2]
                         &   1.18  &  1.20 &   71242 &  1.1985 & -1077 \\
$z^3$I$^o_5$ &  68829.56 &   0.81  &  0.83 &   71767 &  0.8396 &  -246 \\

\hline \hline
\end{tabular}
\footnotetext[1]{Ref.~\cite{Sugar}}
\footnotetext[2]{States observed in quasar absorption spectra, Ref.~\cite{cobalt}}
\end{table}

\begin{figure}
\centering
\epsfig{figure=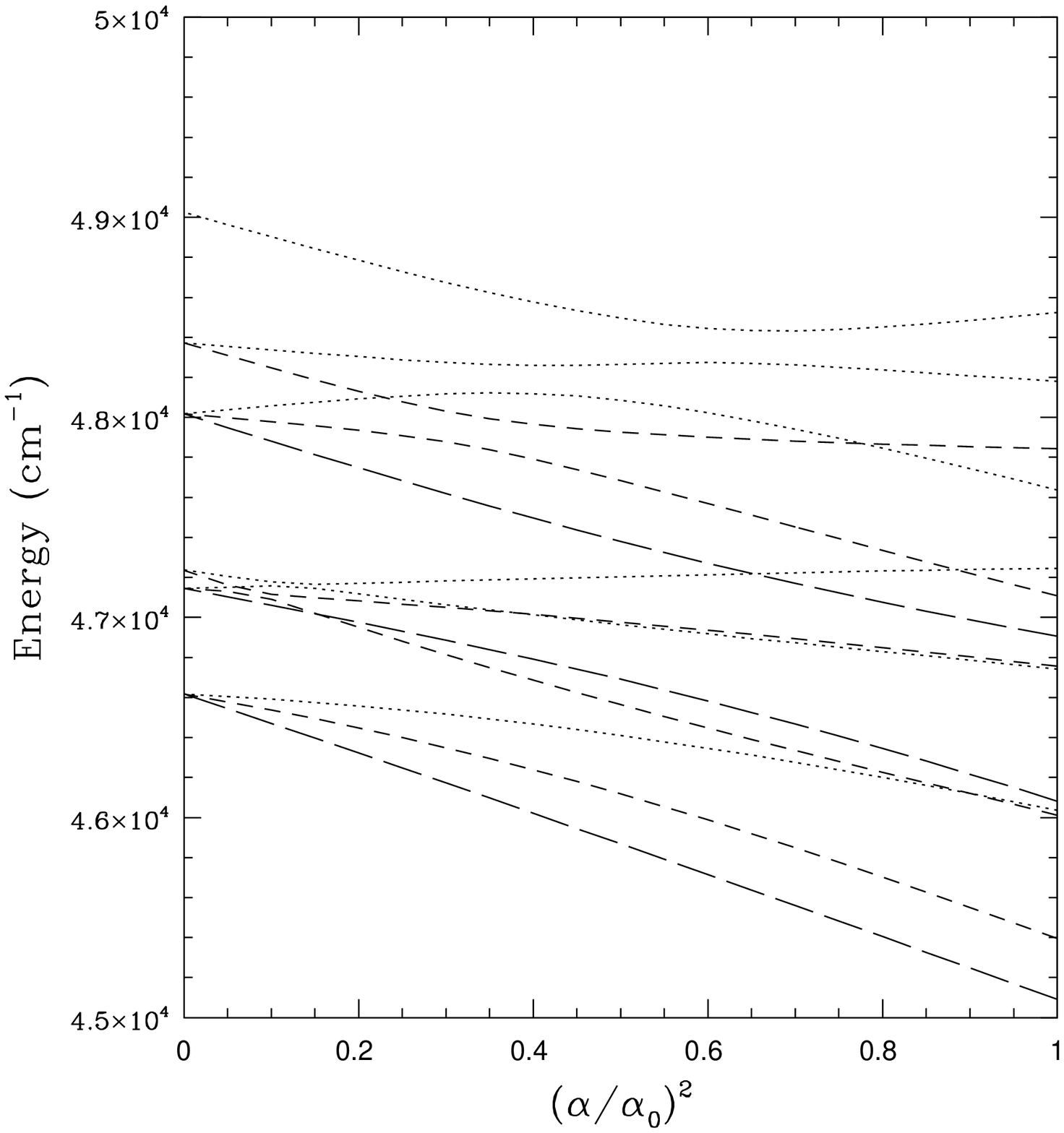,scale=0.45}
\caption{Energy levels of cobalt between 45000 cm$^{-1}$ and 50000 cm$^{-1}$
as functions of $(\alpha/\alpha_0)^2$. Dotted line: $J=3$,
short dash line: $J=4$, long dash line: $J=5$}
\label{co1-fig}
\end{figure}

\begin{figure}
\centering
\epsfig{figure=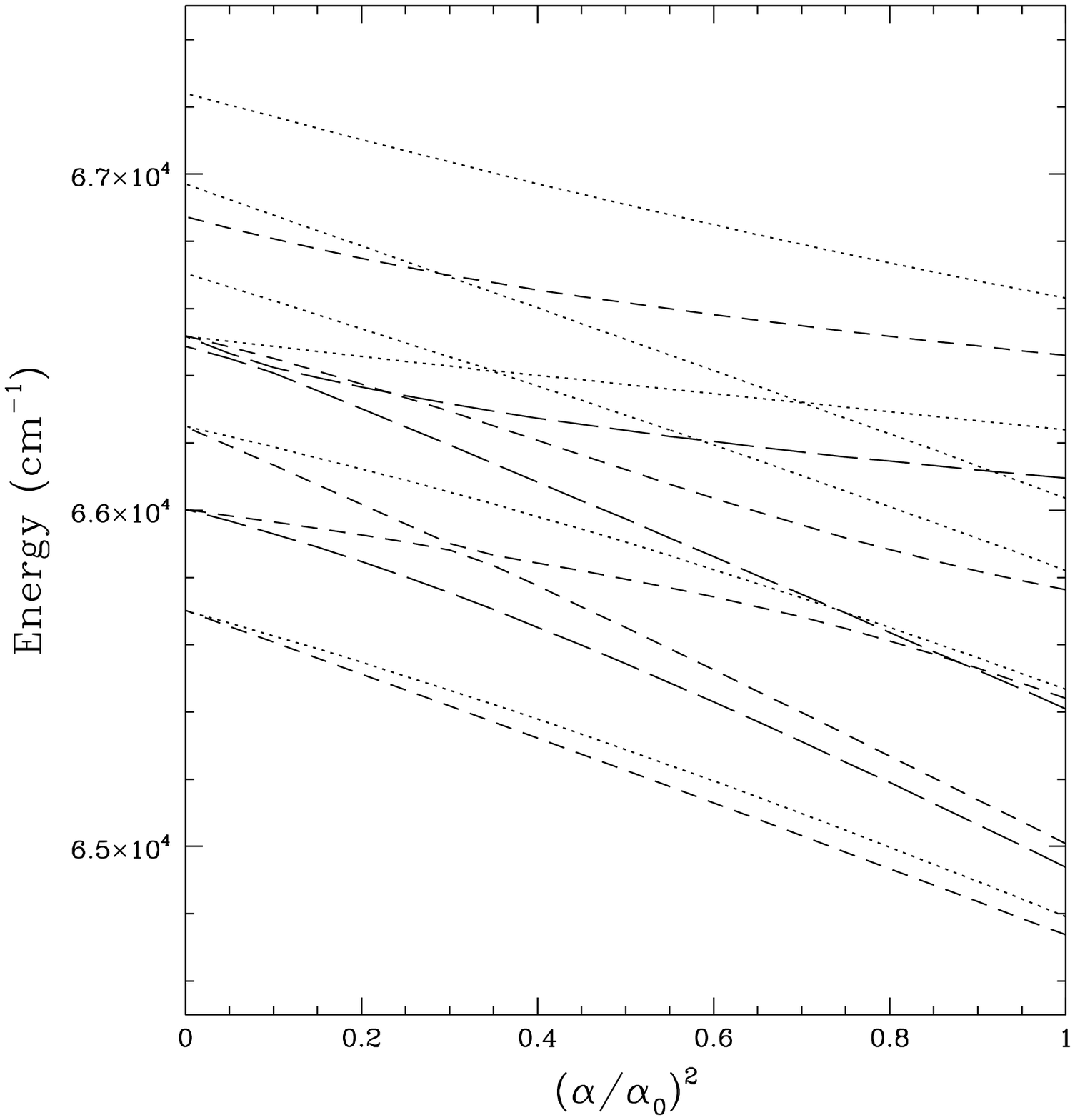,scale=0.45}
\caption{Energy levels of cobalt between 64500 cm$^{-1}$ and 67500 cm$^{-1}$
as functions of $(\alpha/\alpha_0)^2$. Dotted line: $J=3$,
short dash line: $J=4$, long dash line: $J=5$}
\label{co2-fig}
\end{figure}

\begin{figure}
\centering
\epsfig{figure=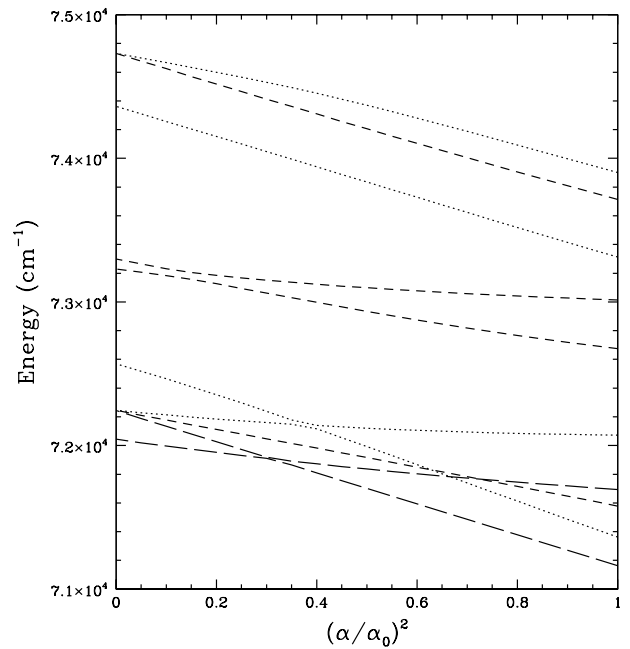,scale=0.45}
\caption{Energy levels of cobalt between 71000 cm$^{-1}$ and 75000 cm$^{-1}$
as functions of $(\alpha/\alpha_0)^2$. Dotted line: $J=3$,
short dash line: $J=4$, long dash line: $J=5$}
\label{co3-fig}
\end{figure}

The results of calculations are presented in Table~\ref{Tab:q}. We
include experimental and theoretical energy levels, $g$-factors and
$q$-coefficients. 
Non-relativistic $g$-factors are given by
\begin{equation}
  g_{NR} = 1 + \frac{J(J+1) - L(L+1) + S(S+1)}{2J(J+1)},
\label{eq:gnr}
\end{equation}
where $J$ is total momentum of the atom, $L$ is its angular momentum
and $S$ is spin.
The $g$-factors are useful for identification of states.
As can be seen from Table~\ref{Tab:q} experimental $g$-factor are close
to the non-relativistic values given by (\ref{eq:gnr}). This
justifies non-relativistic notations for the states and allows to
group them into fine structure multiplets. This is why $g$-factors 
are often more important then energies for the identification of the
states. An interesting example of this kind will be discussed below.

Meanwhile we would like to stress that the accuracy for the calculated
$g$-factors can also serve as an indicator on what kind of accuracy
can be expected for the $q$-coefficients.
This is because $g$-factors are sensitive to configuration
mixing in a very similar way as the $q$-coefficients~\cite{Dzuba02}. 
They can even be used to tune configuration mixing to reproduce 
experimental $g$-factors and therefore improve the accuracy for 
the $q$-coefficients~\cite{Dzuba02}. 
Corresponding procedure can be used when the 
energies of two states with the same total momentum $J$ come very
close to each other in the vicinity of the physical value of
$\alpha$ indicating possible level pseudo crossing when energies are
considered as functions of the ratio $(\alpha/\alpha_0)^2$.
Figures \ref{co1-fig},\ref{co2-fig}
and \ref{co3-fig} show energy levels of Co~II as functions of the fine
structure constant from its non-relativistic limit $\alpha=0$ to the
physical value $\alpha = \alpha_0 = 1/137.036$. One can see that there
are many level crossings but most of them are at safe distance from
the physical point $\alpha = \alpha_0$.  

An interesting example is the mixing of three levels of total 
momentum $J=3$: $E$=  50036.55 cm$^{-1}$,
$E$=50381.86 cm$^{-1}$ and $E$=51512.41 cm$^{-1}$. Fig.~\ref{co1-fig}
shows that these three levels cross each other at about 
$(\alpha/\alpha_0)^2=0.5$. This probably means that mixing should have 
little effect on $g$-factors or $q$-coefficients at $(\alpha/\alpha_0)^2=1$.
However, calculations do not reproduce correct order of states on the
energy scale. Here we have to use comparison between calculated,
experimental and non-relativistic $g$-factors to decide which fine
structure multiplet a calculated state should go to.
 
In spite of the fact that calculations does not always reproduce
correct order of states 
the results for energies and $q$-coefficients are stable with respect
of variation of the details of the calculation procedure.
We estimate that the absolute uncertainty in values of $q$
for all states is 
about the same and of the order of 200 cm$^{-1}$. Therefore, relative 
uncertainty for smaller $q$ is higher. This is because all states belong
to the same $3d^74p$ configuration and small values of $q$ are due to
repulsion from lower states of the same total momentum $J$. 
Variation in the strength of this repulsion or in any other details of
the calculations are likely to change the values of the $q$-coefficients 
for both interacting states by about the same amount~\cite{Dzuba02}.  
Note however that in cases of poor accuracy for $g$-factors and wrong
order of states the uncertainty might be slightly higher than 200 cm$^{-1}$.

\section*{Acknowledgments}

The authors are grateful to J.K. Webb for brining the cobalt data to
their attention and for stimulating discussions.
The work was funded in part by the Australian Research Council and 
Marsden grant.

\end{document}